\pdfoutput=1
\documentclass[aps,prl,reprint,groupedaddress,twocolumn,superscriptaddress]{revtex4-1}

\usepackage{graphicx}
\usepackage{hyperref}
\usepackage{amsmath}
\begin{document}

\title{Mass-selective removal of ions from Paul traps using parametric excitation}
\author{Julian Schmidt}
\address{Albert-Ludwigs-Universit\"{a}t Freiburg, Physikalisches Institut, Hermann-Herder-Stra{\ss}e 3, 79104 Freiburg, Germany}
\address{Laboratoire Kastler Brossel, Sorbonne Universit\'{e}, CNRS, ENS-PSL Research University, Coll\`{e}ge de France, 4 place Jussieu, Paris, France}
\address{National Institute of Standards and Technology, Boulder, CO, USA}
\author{Daniel H\"{o}nig}
	\address{Albert-Ludwigs-Universit\"{a}t Freiburg, Physikalisches Institut, Hermann-Herder-Stra{\ss}e 3, 79104 Freiburg, Germany}
\author{Pascal Weckesser}
\address{Albert-Ludwigs-Universit\"{a}t Freiburg, Physikalisches Institut, Hermann-Herder-Stra{\ss}e 3, 79104 Freiburg, Germany}
\author{Fabian Thielemann}
\address{Albert-Ludwigs-Universit\"{a}t Freiburg, Physikalisches Institut, Hermann-Herder-Stra{\ss}e 3, 79104 Freiburg, Germany}
\author{Tobias Schaetz}
\address{Albert-Ludwigs-Universit\"{a}t Freiburg, Physikalisches Institut, Hermann-Herder-Stra{\ss}e 3, 79104 Freiburg, Germany}
\author{Leon Karpa}
\address{Albert-Ludwigs-Universit\"{a}t Freiburg, Physikalisches Institut, Hermann-Herder-Stra{\ss}e 3, 79104 Freiburg, Germany}
\date{\today}

\begin{abstract}
We study a method for mass-selective removal of ions from a Paul trap by parametric excitation. This can be achieved by applying an oscillating electric quadrupole field at twice the secular frequency $\omega_{\text{sec}}$ using pairs of opposing electrodes.
While excitation near the resonance with the frequency $\omega_{\text{sec}}$ only leads to a linear increase of the amplitude with excitation duration, parametric excitation near $2\, \omega_{\text{sec}}$ results in an exponential increase of the amplitude.
This enables efficient removal of ions from the trap with modest excitation voltages and narrow bandwidth, therefore substantially reducing the disturbance of ions with other charge-to-mass ratios.
We numerically study and compare the mass selectivity of the two methods.
In addition, we experimentally show that the barium isotopes with 136 and 137 nucleons can be removed from small ion crystals and ejected out of the trap while keeping $^{138}\text{Ba}^{+}$ ions Doppler cooled, corresponding to a mass selectivity of better than $\Delta m / m = 1/138$.
This method can be widely applied to ion trapping experiments without major modifications, since it only requires modulating the potential of the ion trap.
\end{abstract}

\maketitle

\section{Introduction}
Radiofrequency (rf) Paul traps \cite{Paul1990,Dehmelt1990,Wineland2013} can store ions for a wide range of charge-to-mass ratios $Q/m$ and thus represent a versatile and powerful tool with applications such as quantum information processing \cite{Wineland2011,Wineland2009} or cold chemistry \cite{Haerter2014,Tomza2019}. Many experiments involve trapping one or two atomic or molecular ion species.  Over the course of the experiment, other (parasitic) isotopes or ion species may appear in the trap, e.g. due to chemical reactions of the ion of interest with background gas particles, and create a disturbance \cite{Schmidt2020}. The parasitic ions can be embedded into the ion crystal after sympathetic cooling \cite{Haerter2013b} or remain on large orbits \cite{Guggemos2015}.

In principle, it is possible to operate any quadrupole mass filter (including rf ion traps) in a parameter regime for which the trajectories of the desired species are stable and those of other species are unstable due to their specific charge-to-mass ratios. The dynamics are given by the solutions of the Mathieu equation and regions of stability can be described with the $a$ and $q$ parameters (see below) which are related to the electrostatic and radiofrequency voltages \cite{Paul1990}. 
Due to technical limitations, it can be challenging to reach such regions of instability in an experiment. Other means of removing ions may then be advantageous, such as resonant excitation \cite{Fulford1980,Londry2003}.

By applying an oscillating voltage to one of the trap electrodes, the near-harmonic motion of the ions in the effective potential can be resonantly excited by displacing the ion from its equilibrium position \cite{McCormick2019}. This method, also known as ``tickle'', will be referred to as \emph{displacement driving} in the following. Since the secular frequency $\omega_{\text{sec}}$ in the effective radial potential is proportional to $Q/m$, this method is also suitable to selectively remove ions from the trap \cite{Fulford1980} with a $Q/m$ resolution depending on the driving duration \cite{Goeringer1992}.

Parametric resonances offer another possibility to efficiently drive a mechanical oscillator \cite{Landau1976}. Ion traps are ideally suited to study  parametric excitation due to the high degree of harmonicity of the typical potentials \cite{Yu1993, Alheit1997,Zhao2002}. Applications range from ion removal with higher $Q/m$ resolution \cite{Vedel1990,Williams1991,Haerter2013b} to excess micromotion compensation \cite{Ibaraki2011,Tanaka2012}. Very recently, experiments in ion traps have exploited parametric excitation to create squeezed motional states for quantum control in both quantum sensing and experimental quantum simulations \cite{Heinzen1990,Burd2019,Wittemer2019,Kiefer2019,Wittemer2020}.

Here, we review and discuss the dynamics of a single ion in a Paul trap under the influence of parametric excitation and experimentally use this method for ion removal. We first perform numerical simulations to compare the ion removal efficiency achieved by either parametric excitation or resonant displacement driving. We find that parametric excitation can remove ions within milliseconds (or a few hundred oscillation cycles) with a $Q/m$ resolution of better than $ 10^{-3} $ and vanishing off-resonant excitation. In addition, we study the case in which an ion of mass $m_2$ should be removed while it is sympathetically cooled by a co-trapped, laser cooled ion of different mass $m_1$ which should remain in the trap. Finally, we discuss a first characterization of the experimental performance of the method and compare the results with our numerical calculations. 

The method is flexible, as it only requires modulating the potential of the trap (in our case via the rf driving amplitude), representing a minor modification for typical ion trapping experiments. It can be used for any set of $a$ and $q$ parameters and does not require tuning the trap to a higher-order (e.g. octupolar) instability \cite{Franzen1994,Drakoudis2006}.
This method may be applicable to chemistry experiments which study reactions of ions and a (neutral) atomic ensemble \cite{Schmidt2020,Haerter2012,Chen2014,Greenberg2018,Gingell2010}, but also to precision measurements where the spectroscopy ions are sympathetically cooled by an ion species amenable to laser cooling \cite{Schmoeger2015,Biesheuvel2017,Alighanbari2018,Chou2017}. It could be extended to ions in Penning traps or storage rings.

\section{Theoretical model}\label{sec:theory}

We consider a linear rf ion trap oriented along the $z$-axis. We discuss the experimental realization later and refer to section~\ref{sec:exp} and Fig.~\ref{fig:setup}. Two rf electrodes provide dynamical radial confinement while two (segmented) dc electrodes control the electrostatic axial confinement along $z$ as well as radial electric fields. The motion of the ion in the Paul trap in the radial $x-y$-plane, here exemplarily shown for $x$, is described by the Mathieu equation \cite{Leibfried2003}:
\begin{equation}
 \frac{d^2x}{dt^2} + \frac{Ze}{m} \left[ U_{\text{dc}} \alpha_x + U_{\text{rf}} \alpha_x^{\prime} \cos(\Omega t)\right] x = 0 \label{eq:Mathieu}\;
\end{equation}

where $Ze=Q$ is the charge of the ion, $m$ its mass, $U_\text{dc}$ and $U_\text{rf}$ are the voltages applied to the dc and rf electrodes and $\Omega$ is the angular frequency of the rf driving field. $\alpha$ and $\alpha^{\prime}$ are geometry factors with unit m$^{-2}$ determined by the trap dimensions. The motion along the $y$ and $z$ directions can be written analogously. The Mathieu equation can be rewritten in terms of the $a$ and $q$ parameters, which we will use in the following:
\begin{equation}
a_x = \frac{4Ze U_{\text{dc}} \alpha}{ m \Omega^2}\; , \quad q_x =\frac{2Ze U_{\text{rf}} \alpha^\prime}{m\Omega^2} \;.
\end{equation}

In the limit $a_x \ll 1$, $q_x \ll 1$, the solution of the Mathieu equation can be approximated as a secular motion at frequency $\omega_{x,\text{sec}}$ (along $x$) superimposed with a fast micromotion at $\Omega$ \cite{Leibfried2003}. In this so-called pseudopotential approximation, $\omega_{x,\text{sec}} \approx (\Omega/2) \sqrt{a_x + q_x^2/2}$ \cite{Paul1990}.

Within the pseudopotential approximation, in which the micromotion is neglected, a modulation of $U_{\textrm{rf}}$ at the frequency $\omega_{\textrm{mod}}$ can again be modelled by the Mathieu equation. Similarly to the undriven ion trap, the parametrically driven harmonic oscillator exhibits regions of stability and instability, as has been previously shown and discussed in detail \cite{Zhao2002}.

In order to study the dynamics of the problem and to study the experimental relevance of the method for various scenarios, we numerically solve the Mathieu equation, Eq. \ref{eq:Mathieu}, after adding a parametric drive.
The equation of motion is
\begin{equation}
 \frac{d^2x}{dt^2} + \frac{1}{2} \Omega^2 x \left[\frac{a_x}{2} + q_x \left(1 + A_{p} \sin\left({\omega_{\textrm{mod}} t}\right) \right) \cos\left({\Omega t}\right) \right] = 0 \; \label{eq:secondmathieu}.
\end{equation}

We solve this equation in one dimension for our experimental parameters, $\Omega = 2\pi \times 1.4\,\textrm{MHz}$, $a = 0.001$, $q=0.25$, $m=m_{\text{Ba}} = 138~\text{amu}$ (throughout this paper, $m_{\text{Ba}}$ refers to the mass of $^{138}\text{Ba}^{+}$). We tune $\omega_{\textrm{mod}} \approx 2 \omega_{x,\text{sec}}$ with an amplitude (modulation depth) up to $A_{p} = 6 \times 10^{-3}$. On resonance, the parametrically driven oscillator becomes metastable and starts a coherent oscillation for an initial, nonzero displacement from the origin \cite{Landau1976}. As initial conditions, we therefore choose the position $x_{\text{ini}}=1 \,\mu\text{m}$ and momentum $p_{\textrm{ini}} = 0$. Our parameters correspond to a secular frequency of $\omega_{\text{sec}} \approx 2\pi \times 127 \, \textrm{kHz}$ and an initial potential energy of about $k_{\textrm{B}} \times 5 \,\textrm{mK}$ (and zero kinetic energy).

\begin{figure}
 \resizebox{1\hsize}{!}{\includegraphics*{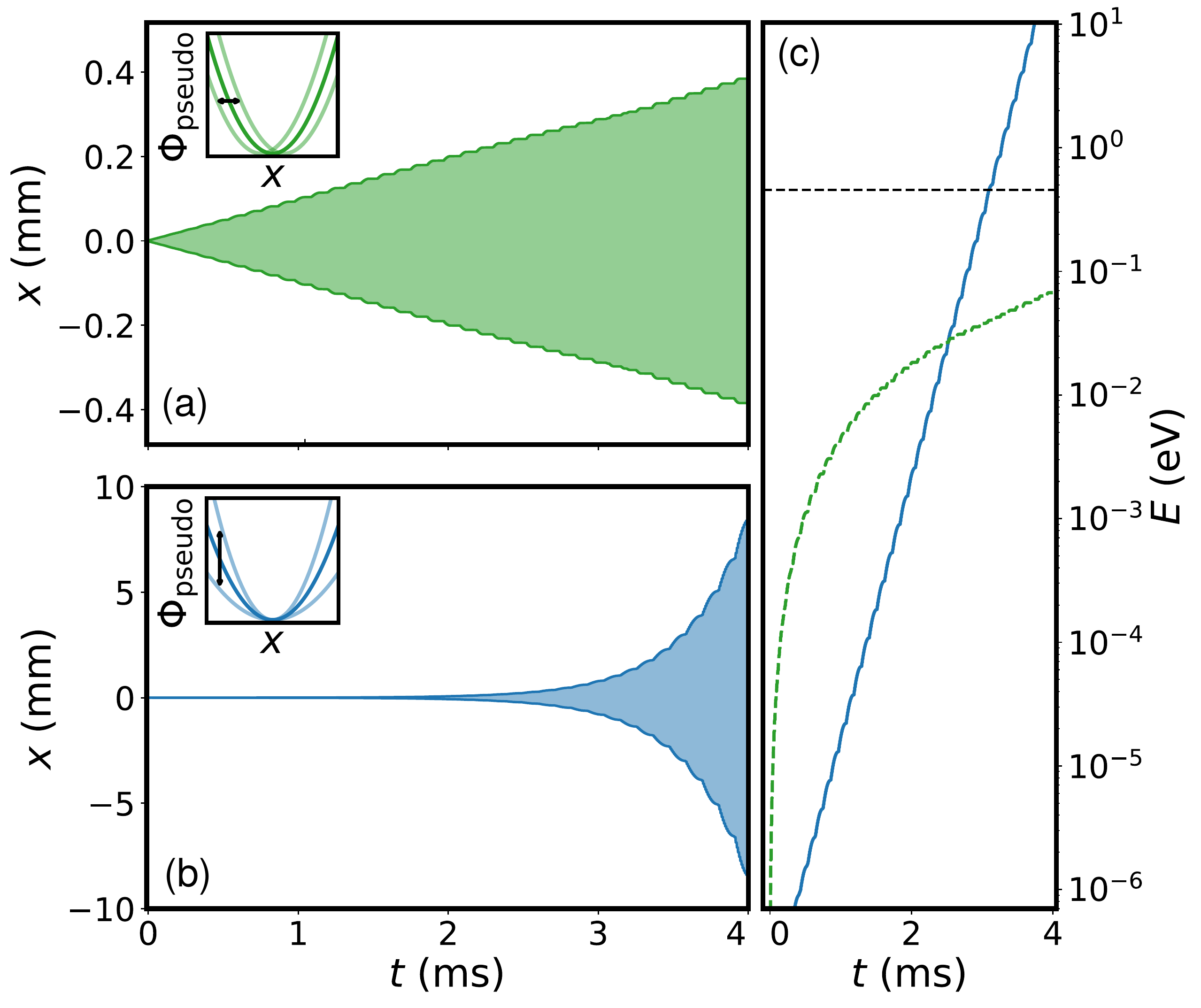}}
 \caption{Numerical calculation of the trajectory of an ion in a Paul trap under (a) the influence of an oscillating linear electric field at the resonance frequency (displacement driving, modulates the displacement of the potential) and (b) an oscillating quadrupole electric field applied to the rf electrodes (parametric driving, modulates the potential curvature). The plots depict the envelope of the oscillation and show the linear versus exponential growth of the amplitudes. In (c), the total (kinetic and potential) energy of the ion is shown on a logarithmic scale in the case of displacement (dashed curved line) and parametric driving (solid line). The dashed horizontal line corresponds to a motional amplitude of $1 \, \text{mm}$ (zero to peak), at which it is in general assumed that an ion is lost from the trap. For displacement driving, we choose an electric field amplitude of $0.2\,\text{V/m}$ and for parametric driving, a modulation depth of $6\times 10^{-3}$, both at a secular frequency of about $\omega_{x,\text{sec}} = 2\pi \times 127\,\text{kHz}$. \label{fig:typical_trap}}
\end{figure}

Typical trajectories are shown in Fig. \ref{fig:typical_trap}. Since the parametric modulation term $q_x A_p \sin \left({\omega_{\text{mod}}t} \right) $ in Eq. \ref{eq:secondmathieu} is proportional to the oscillation amplitude $x(t)$ at any time, the energy in the system increases exponentially \cite{Landau1976,Heinzen1990,Yu1993}, see Fig. \ref{fig:typical_trap} (c). The rate of energy gain is proportional to $x_{\text{ini}}$, $A_{p}$ and $q_x$. 
We compare this result with resonant displacement driving in Fig. \ref{fig:typical_trap} (a) and (c), where the amplitude increases linearly and the energy increases quadratically. In the displacement driving case, the motion in $x$ ($y$ analogous) is described by
\begin{equation}
 \frac{d^2x}{dt^2} + \frac{1}{2} \Omega^2 x \left[\frac{a_x}{2} + q_x \cos\left({\Omega t} \right) \right] + \frac{Ze}{m} A_{d} \sin\left(\omega_{mod} t \right) = 0\; .
\end{equation}
In our experiment, a typical value for the amplitude is given by $A_{d} = 1\, \textrm{V/m}$, corresponding a voltage amplitude of $10 \, \textrm{mV}$ (all amplitudes given as zero to peak) on a given dc electrode, in agreement with our segmented linear trap geometry with an ion-electrode distance of 9~mm, see section \ref{sec:exp}.

The mass resolution is determined by the response of a single ion to the excitation as a function of the ratio $m/(Ze)$, with $Z = 1$ in our case. To compare the mass resolution in both the parametric and the displacement driving cases, we solve the equations of motion for the parameters given above. The modulation frequency $\omega_{\textrm{mod}}$ (chosen to excite $^{138}\text{Ba}^{+}$) is constant while $m$ is varied in discrete steps within the range $m_{\text{Ba}} \pm 3\%$. We choose the duration $t_{\text{exc}}$ and amplitude of the excitation for both parametric and displacement driving such that on resonance, the same oscillation amplitude of the ion is reached after $t_{\text{exc}}$ (intersection of the two curves in Fig. \ref{fig:typical_trap} (c)).

\begin{figure}
 \includegraphics[width=0.45\textwidth]{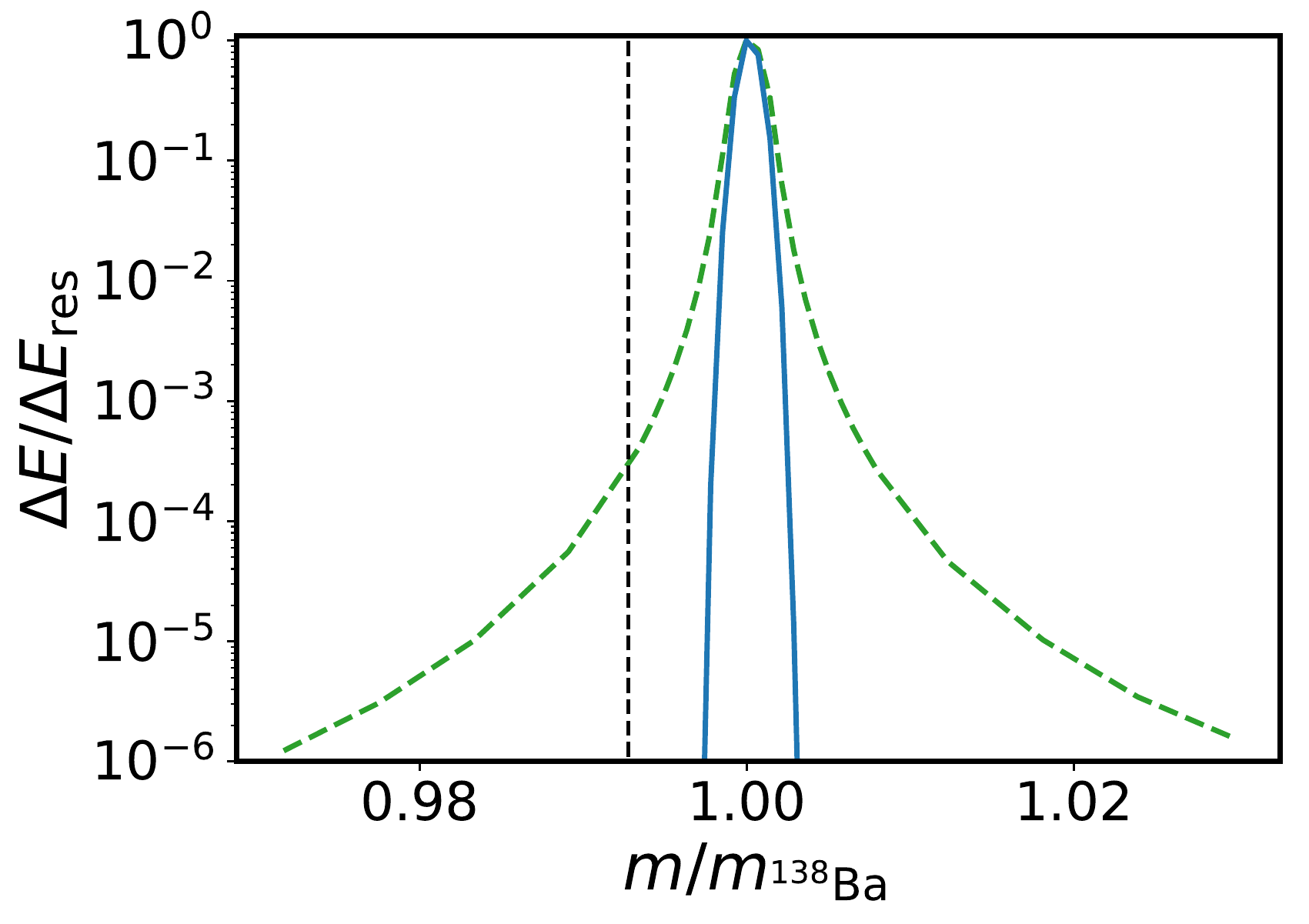}
 \caption{Mass dependence of the energy gain using parametric excitation compared to displacement driving. We simulate the dynamics of an ion for different $m/m_{\text{Ba}}$ in both cases while keeping the excitation frequency constant. At $t_{\text{exc}}$, we calculate the energy of the ion, which is plotted on a logarithmic scale. The driving amplitudes have been chosen as in Fig. \ref{fig:typical_trap}, and in order to ensure comparability, $t_{\text{exc}}=2.54 \,\text{ms}$ corresponds to the point in Fig. \ref{fig:typical_trap}(c) where the energy curves for parametric and displacement driving intersect. Here, the parametric resonance is narrower than the displacement driving resonance by a factor of about 1.6 (full width at half maximum). More importantly, the tail of the parametric driving case falls off as a Gaussian, while that of the displacement driving resonance behaves as a Lorentzian function. Parametric driving thus acts as a knife for the ions close to resonance, allowing us to anticipate the separation of two ions with a mass ratio of $137/138$ within milliseconds, as depicted by the dashed vertical line.
  \label{fig:mass_res}}
\end{figure}

The results are shown in Fig. \ref{fig:mass_res} and clearly emphasize the improved mass resolution with parametric excitation. While the full width at half maximum (FWHM) of the response to the parametric driving is only smaller by a factor of 1.6, the striking difference is the response when further detuned from the resonance. The response to displacement driving has a Lorentzian shape and falls off with a power law, while the response to parametric driving is Gaussian and falls off exponentially for $m\neq {138}\, \textrm{amu}$ and allows for an improved mass resolution. This is equivalent to an improvement of the frequency resolution. For a relative mass difference \linebreak $|m - m_{\text{Ba}}| / m_{\text{Ba}} > \pm 0.2\%$, parametric driving already suppresses the excitation by 40~dB  compared to displacement driving.

\section{Separation of ions with different masses}

We now study the case of two co-trapped ions 1 and 2 which form a crystal and are strongly coupled. Ion 1 has mass $m_1$ and is to remain in the trap while ion 2 with mass $m_2$ is a parasitic ion which is to be removed from the trap. We numerically show that this can be achieved by cooling the ion 1 while parametrically driving ion 2. In this case, the crystal formed by the two ions has to melt in order to decouple the motion of the two ions.

We tested that performing the calculation with the full dynamics including $\Omega$ or within the pseudopotential approximation does not significantly change the dynamics of the ions in the simulation. In this section, we therefore use the pseudopotential approximation to speed up the calculation while extending it to three dimensions.

Along $x$, each of the two ions would individually oscillate at a secular frequency $\omega_{x,\text{sec} 1}$ and $\omega_{x,\text{sec} 2}$, where the frequencies depend on $m_1$ and $m_2$ (analogous for $y$, $z$).
Due to the strong coupling of the ions, their motion can be described in the normal mode picture with $3 N$ ($N=2$) oscillation frequencies. By exposing ions located near their equilibrium positions to homogeneous electric fields, only the common modes can be excited. However, as the oscillation amplitudes grow, the coupling between the ions becomes weaker until the normal mode description breaks down.

For a two-ion crystal with $m_1 = m_2$, the motion in $x$ can be decomposed into the center-of-mass mode $\omega_{x,\text{com}} = \omega_{x,\text{sec}}$ and the so-called rocking mode
\begin{equation}
 \omega_{x,\text{rock}} = \sqrt{\omega_{x,\text{com}}^2-\omega_{z,\text{com}}^2} \, , \label{eq:rocking}
\end{equation}
where $\omega_{z,\text{com}} = \omega_{z,\text{sec}}$ is the center-of-mass frequency along the trap axis $z$.
Generally, if $m_1 \neq m_2$, the radial modes have a different structure and the center-of-mass frequency is not equal to either $\omega_{x,\text{sec} 1}$ or $\omega_{x,\text{sec} 2}$. Parametric driving at the frequency of the unperturbed ion 1 ($\omega_{\text{mod}} = 2 \omega_{x,1}$), is then not resonant with the common motion of the ions. The individual detuning from resonance, which is related to the excitation efficiency, depends on $m_1/m_2$. In addition, the sympathetic cooling within the Coulomb crystal is also affected by the mass ratio (more efficient for mass ratios closer to 1), leading to a non-trivial behaviour of the excitation as long as the ions remain close to their coupled equilibrium positions.
For our experimental parameters, the residual off-resonant excitation is sufficient to amplify the relative motion of the ions. Once the relative motion of the ions is sufficiently large, the ions are effectively decoupled and ion 1 is further excited parametrically as in the single ion case.
The shift of the frequencies of the radial normal modes also depends on the axial confinement, as is apparent from Eq. \ref{eq:rocking}. In our case, this shift is small ($\approx 1\%$) as $\omega_{x,\text{com}} \gg \omega_{z,\text{com}}$, which is beneficial for effectively decoupling the ions in the parametric excitation scheme.

\begin{figure}
 \includegraphics[width=0.5\textwidth]{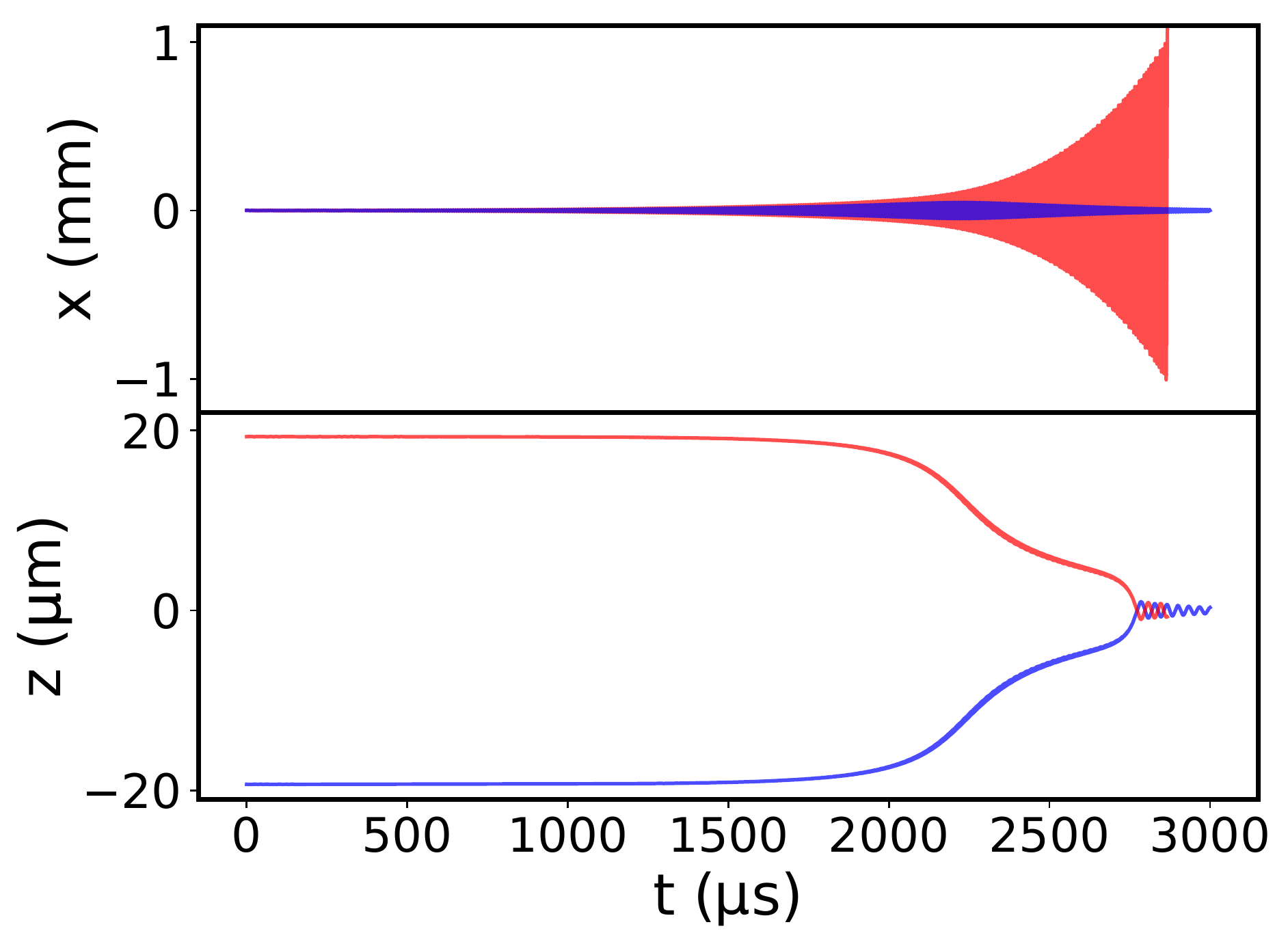}
 \caption{Calculated trajectories of two ions with different masses in the rf trap. Here, the ions are considered to be $^{137}\text{Ba}^{+}$ (red curve) and $^{138}\text{Ba}^{+}$ (blue curve). The motion of the $^{138}\text{Ba}^{+}$ ion is damped by laser cooling and sympathetically cools the $^{137}\text{Ba}^{+}$ ion through the mutual Coulomb interaction. We set the common initial displacement in the $x$ direction to $1\,\mu\text{m}$. At $t=0$, a parametric excitation at twice the resonance frequency of a single $^{137}\text{Ba}^{+}$ ion is turned on while the $^{138}\text{Ba}^{+}$ ion is still directly laser cooled. The $x$ (radial) and $z$ (axial) positions of the ions are plotted as a function of time. Initially, the ions are aligned along the $z$ axis within a common ion crystal, with their Coulomb interaction in equilibrium with the axial electrostatic potential. The ions then begin oscillating in the $x$ direction in common, until the motion of the $^{138}\text{Ba}^{+}$ ion is sufficiently decoupled and only the amplitude of the $^{137}\text{Ba}^{+}$ ion keeps growing. Here, this critical point is reached after about 2ms and the $^{138}\text{Ba}^{+}$ is cooled back to the trap center. The amplitude of the $^{137}\text{Ba}^{+}$ reaches $1\,\text{mm}$ at $t \approx 2.8\,\text{ms}$, where it is assumed to leave the trap.
  \label{fig:ion_sep}}
\end{figure}

The simulation results for two ions with an additional damping term for ion 1 are shown in Fig. \ref{fig:ion_sep}. The Coulomb crystal is initially radially displaced by $1\,\mu\text{m}$. Here, we use a higher modulation amplitude for parametric driving than before to reduce the computation time followed by a qualitative analysis only. When comparing the trajectories of the two ions, we observe that the motion of the two ions is initially driven in common, that is, they still form a Coulomb crystal. Ion 1 (blue) is cooled and damps the motion of ion 2 (red). At a critical amplitude, ion 2 ``breaks free'' and the motion is decoupled from that of ion 1. The amplitude of ion 2 then increases exponentially as in the single ion case while ion 1 is again efficiently Doppler cooled. 


\section{Comparison with experimental results}\label{sec:exp}

\begin{figure}
 \includegraphics[width=0.4\textwidth]{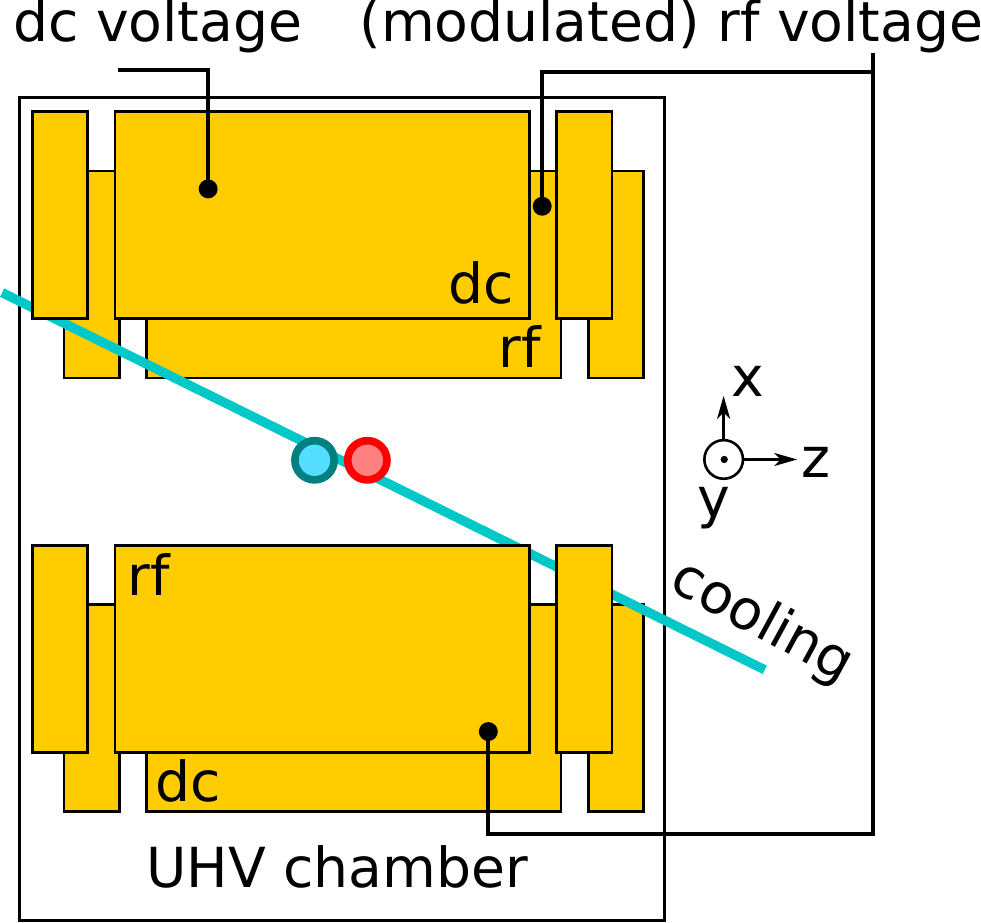}
 \caption{Sketch of the experimental setup. $^{138}\text{Ba}^{+}$ ions (blue circle) are trapped and laser cooled near the center of a segmented linear rf trap. Other isotopes of barium (red circle) are also produced during photoionization and trapped. In addition, we can produce magneto-optically and all-optically trapped clouds of $^{87}$Rb, which sometimes leads to the production of Rb$^{+}$ and Rb$_2^{+}$ ions \cite{Haerter2013b,Schmidt2020}. These ions can be crystallized (strongly coupled to the laser cooled ions via Coulomb interaction) or in a plasma (weakly coupled). In order to remove all ions but $^{138}\text{Ba}^{+}$, we modulate the rf drive of our Paul trap at twice the radial secular frequency of these ions through the helical resonator. We can also apply an oscillating displacement via the dc electrodes for determining the trapping frequencies. 
  \label{fig:setup}}
\end{figure}

\begin{figure}
 \includegraphics[width=0.5\textwidth]{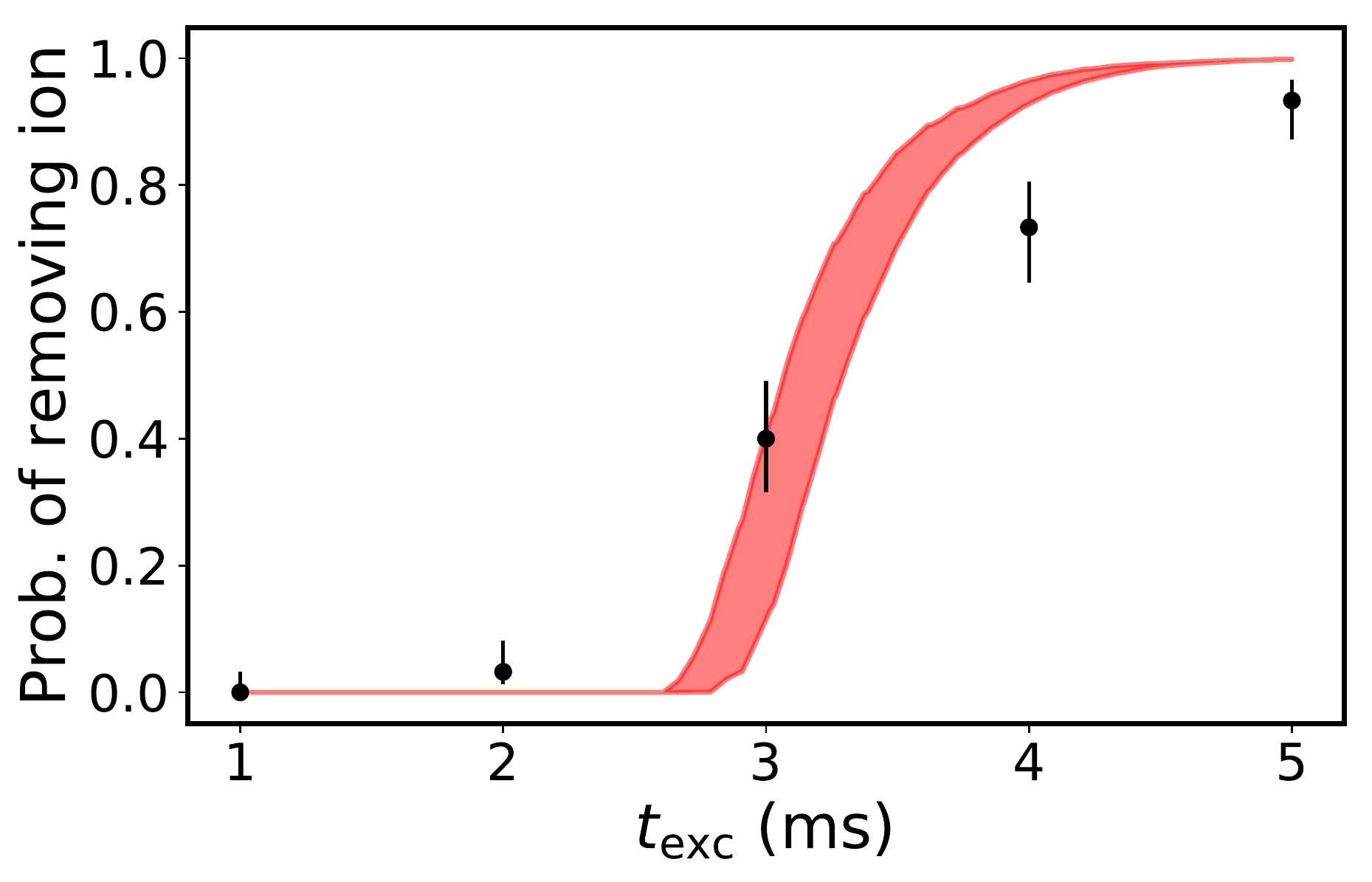}
 \caption{Measurement of the timescale for removing a single ion via parametric excitation. After Doppler cooling, at $t=0$, the cooling and repumping lasers are turned off and the rf trap drive is modulated to parametrically excite the ion at $2\omega_{\text{sec}}$. After $t_{\text{exc}}$, the parametric excitation is turned off and the cooling and repumping lasers are turned on again. Then, the CCD camera signal is monitored to observe whether the ion is recooled into the trap. If, after 1~s, no fluorescence signal can be observed, we consider the removal attempt from the trap as successful. 
 After loading a new ion if necessary, the experiment is repeated until each $t_{\text{exc}}$ has been repeated about 30 times. Finally, the removal probability is shown in dependence on the excitation duration $t_{\text{exc}}$. Due to the finite temperature of the ion, the initial conditions for position and momentum vary according to the Boltzmann distribution and lead to a smoothed transition from 0\% to 100\% removal probability. The shaded area is an estimate of the removal probability of the ion after $t_{\text{exc}}$ based on a numerical calculation of its trajectory under the influence of parametric excitation and assuming an initial temperature of $T \approx 360\, \mu$K corresponding to the Doppler limit. The ion is considered to be removed when it reaches a distance of 1~mm from the trap center.
  \label{fig:exp_res}}
\end{figure}

We next perform an experimental investigation of parametric driving as a tool for removing ions from a Paul trap.
Our experimental setup \cite{Lambrecht2017,Schmidt2018,Schmidt2020} shown in Fig. \ref{fig:setup} consists of a linear segmented Paul trap (with ion-electrode distance of 9 mm) under ultra-high vacuum (UHV). The rf source is a signal synthesizer operated at a frequency $\Omega = 2\pi \times 1.4\,\text{MHz}$. A helical resonator with a quality factor of 40 enhances $U_{\text{rf}}$ up to 2000~V and is connected to two diagonally opposed electrodes of the Paul trap via a standard vacuum high-frequency feedthrough. The other two segmented electrodes are grounded with respect to the rf electrodes and can be offset by small dc fields to compensate stray fields \cite{Huber2014} and to provide axial confinement of typically $\omega_{z,\text{sec}} = 2\pi \times 12\,\text{kHz}$. $^{138}\text{Ba}$ atoms (and other isotopes) are emitted from an oven, photoionized with a resonant 553~nm and an additional 405~nm laser \cite{Leschhorn2012}, and subsequently Doppler cooled to below one millikelvin using lasers at 493~nm and 650~nm for cooling and repumping, respectively. The fluorescence emitted by $^{138}\text{Ba}^+$ during Doppler cooling is imaged by a charge-coupled device (CCD) camera used for detection.

We first experimentally confirm the efficiency of the parametric excitation method, here with a single $^{138}\text{Ba}^+$ ion as a showcase, see Fig. \ref{fig:exp_res}. In order to investigate the undamped dynamics as in Fig. \ref{fig:typical_trap}, we first prepare a single Doppler-cooled $^{138}\text{Ba}^{+}$ ion in the trap. We then turn off the cooling and repumping lasers and turn on parametric excitation for a variable duration $t_{\text{exc}}$. Here, $\omega_{x,\text{sec}} = 2\pi \times(120\pm 1) \, \text{kHz}$ and the corresponding modulation frequency is $\omega_{\text{mod}} = 2\pi \times (240\pm 2) \, \text{kHz}$ with a modulation depth of $8\times 10^{-3}$. We then turn on the cooling and repumping lasers and detect via fluorescence imaging whether the ion has remained in the trap. The probabilistic distribution of the initial position and the velocity after laser cooling translates to a distribution of the duration after which the ion is removed from the trap. This leads to the smooth transition from 0\% removal probability to near-unity removal probability after about 5 ms. The experimental results are in reasonable agreement with the results of our simulation depicted by the red curve. Given that the growth of the motional amplitude is exponential, an accurate description of timescales is challenging.

In practice, using a linear frequency chirp of about 1~kHz and sweep time of 1~ms across the resonance improves the repeatability in our experiment which we attribute to two effects. First, due to the high frequency resolution, small drifts of the secular frequency on the order of 300~Hz require frequent tuning of the modulation frequency. Second, we suspect that anharmonicities of the trap lead to a change in $\omega_{x,\text{sec}}$ at higher oscillation amplitudes. After reaching a critical amplitude, $\omega_{\text{mod}}$ becomes detuned from $\omega_{x,\text{sec}}$ and the modulation no longer efficiently excites the motion. The frequency chirp mitigates both possible issues.

\section{Application to mass separation of mixed-species or mixed-isotope ion crystals and clouds}

As discussed above, various processes can involve loading of contaminating ion species into the trap besides the laser-cooled ion species of interest. For two of such processes, we apply parametric excitation to remove the contaminating ion species.

The photoionization process is not perfectly isotope selective in our experiment due to an angle of about $110^{\circ}$ between the photoionization laser axis and the direction of the hot atomic beam from the oven. This leads to loading of barium isotopes with 136 and 137 nucleons into the trap with a probability of about 10\%. Reactions of electronically excited $^{138}\text{Ba}^{+}$ with background $\text{H}_2$ molecules can also lead to the formation of $\text{BaH}^{+}$ hydride ions \cite{Molhave2000,Kahra2012,Aymar2012}. After sympathetic cooling by $^{138}\text{Ba}^{+}$, the ensemble consisting of $^{138}\text{Ba}^{+}$ and another ion species forms a crystal. Only the $^{138}\text{Ba}^{+}$ ion emits fluorescence from the near-resonant cooling laser and appears as a bright spot on the CCD camera image. The presence of another (dark) ion shifts the $^{138}\text{Ba}^{+}$ ion from its equilibrium position at the center of the trap.

If the fluorescence image indicates that such a mixed crystal is present in the trap, we apply parametric driving on the trap electrodes. We tune the modulation frequency to the resonance of $^{136}\text{Ba}^{+}$ ($^{137}\text{Ba}^{+}$) while laser cooling $^{138}\text{Ba}^{+}$. We observe that the dark ions can be removed from the trap by applying the parametric driving resonant with the secular frequency of $^{136}\text{Ba}^{+}$ ($^{137}\text{Ba}^{+}$). This procedure now takes a few seconds to successfully remove the parasitic ions. We were not able to remove the parasitic ions using displacement driving while the laser-cooled $^{138}\text{Ba}^{+}$ remain in the center of the trap.

Compared to the numerical simulations presented in Fig. \ref{fig:ion_sep}, where it was possible to remove ion 2 within milliseconds while ion 1 remained in the trap, we experimentally observe longer timescales. In our simulations, we found that the behaviour depends on the damping rate, $\omega_{x,\text{sec}}/\omega_{z,\text{sec}}$ and $A_p$. Reaching the critical point at which the parasitic ion is sufficiently decoupled from the laser-cooled ion strongly depends on the initial conditions \cite{Landau1976,Yu1993} and can be challenging. In our experiment, this may be assisted by instantaneous heating effects. Given the observed timescales and our background pressure on the order of $5\times 10^{-10} \, \text{mbar}$, it is possible that background gas collisions initiate the process. The effective continuous sampling of the Boltzmann distribution of the ion while it is being laser cooled may also assist the process.

We have also used this method to remove $\text{Rb}^{+}$ and $\text{Rb}_2^{+}$ ions from our trap \cite{Schmidt2020}. In this case, the ions are typically not embedded into the crystal structure at the center of the trap. Ions on high orbits with large kinetic energy can be removed reliably within tens of milliseconds while only minimally disturbing the $^{138}\text{Ba}^{+}$ in the center of the trap.

\section{Summary and outlook}

We have discussed parametric excitation as a method for mass-selectively removing ions from an rf ion trap. We find that parametric excitation at twice the ion's resonance frequency is a highly efficient and fast method for this purpose. For the high oscillation amplitudes (10\% to 50\% of the ion-electrode distance) required to remove ions from the trap, the mass resolution is significantly enhanced compared to displacement driving at the resonance frequency. Efficiently removing ions with parametric excitation requires precise tuning of the excitation frequency. Small drifts of the trapping frequency can thus limit the performance of the method. While the secular frequencies can be stabilized via rf and dc voltages \cite{Johnson2016}, we used a sweep of the excitation frequency in a 1~kHz window near the resonance. 
Further improvement could be achieved by initially transferring a controlled amount of energy to the ions by applying a resonant displacement driving pulse at $\omega_{x,\text{sec}}$ \cite{Yu1993,McCormick2019} or giving the ions a kick with a voltage pulse.

The amplitude modulation index is limited due to the bandpass filter constituted by the helical resonator. Other schemes to enhance the rf-amplitude which lead to less damping of the modulation sidebands may be advantageous, especially at low trap drive frequencies where higher voltages can be achieved in the pre-amplification. Using even higher harmonics $n\,\omega_{\text{sec}}$ (with even integer $n$) could further improve the frequency resolution \cite{Zhao2002}.
Parametric excitation may be useful when initializing a large ion trap with a single ion, e.g. in the context of shallow optical trapping potentials \cite{Lambrecht2017}.
Due to the recent interest in experiments relying on a multitude of ion species in a single trap, including ions for sympathetic cooling and e.g. molecular ions, we believe that removal of ions by parametric excitation may be a useful tool for many ion trapping groups.

\textbf{Acknowledgements}

This project has received funding from the European Research Council (ERC) under the European Unions Horizon 2020 research and innovation program (Grant No. 648330), and was supported by the Georg H. Endress foundation. J.~S., F.~T., and P.~W. acknowledge support from the DFG within the GRK 2079/1 program. J.~S. acknowledges financial support from the Region \^{I}le-de-France within the framework of DIM SIRTEQ and the Alexander-von-Humboldt foundation. P.~W. gratefully acknowledges financial support from the Studienstiftung des deutschen Volkes. We are indebted to D. Leibfried for stimulating discussions.

\end{document}